\documentclass[a4paper, amsfonts, amssymb, amsmath, reprint, showkeys, nofootinbib, twoside, 10pt]{revtex4-1}
\usepackage[english]{babel}
\usepackage[utf8]{inputenc}
\usepackage[colorinlistoftodos, color=green!40, prependcaption]{todonotes}
\usepackage{amsthm}
\usepackage{mathtools}
\usepackage{physics}
\usepackage{xcolor}
\usepackage{graphicx}
\usepackage{bm}
\usepackage[left=15mm,right=15mm,top=35mm,columnsep=25pt]{geometry} 
\usepackage{adjustbox}
\usepackage{placeins}
\usepackage[labelfont=bf, font=small]{caption}
\usepackage[T1]{fontenc}
\usepackage{lipsum}
\usepackage{csquotes}
\usepackage{layouts}
\usepackage{natbib}
\usepackage[pdftex, pdftitle={Article}, pdfauthor={Author}]{hyperref}
\usepackage[]{lineno}

\newcommand{\ka}{K$\alpha\;$}
\newcommand{\kb}{K$\beta\;$}
\newcommand{\kbt}{K$\beta_{1,\!3}\;$}
\newcommand{\kbp}{K$\beta^\prime\;$}

\newcommand{\fe}[1]{Fe$^{+#1}$}
\begin{document}
%
\title{Detecting Iron Oxidation States in Liquids with the VOXES Bragg Spectrometer}
\author{Simone Manti\textsuperscript{1}}
\author{Marco Miliucci\textsuperscript{1$\dagger$}}
\author{Alessandro Scordo\textsuperscript{1*}}
\author{Roberto Bedogni\textsuperscript{1}} 
\author{Alberto Clozza\textsuperscript{1}}
\author{Mihail Iliescu\textsuperscript{1}}
\author{Gabriel Moskal\textsuperscript{2}} 
\author{Kristian Piscicchia\textsuperscript{1,3}} 
\author{Alessio Porcelli\textsuperscript{3}}
\author{Diana Sirghi\textsuperscript{1,3,4}}
\author{Florin Sirghi\textsuperscript{1}}
\author{Catalina Curceanu\textsuperscript{1}}
\affiliation{}
\affiliation{\textsuperscript{1}INFN, Laboratori Nazionali di Frascati, Via E. Fermi 54, I-00044 Roma, Italy;}
\affiliation{\textsuperscript{2}The M. Smoluchowski Institute of Physics, Jagiellonian University, 30-348 Kraków, Poland}
\affiliation{\textsuperscript{3}Centro Ricerche Enrico Fermi–Museo Storico della Fisica e Centro Studi e Ricerche “Enrico Fermi”, Via Panisperna 89a, I-00184 Roma, Italy}
\affiliation{\textsuperscript{4}IFIN-HH, Institutul National pentru Fizica si Inginerie Nucleara Horia Hulubei, Str. Atomistilor No. 407, P.O. Box MG-6 Bucharest-Magurele, Romania}
\affiliation{$^*\!\!\!$ Correspondence: alessandro.scordo@lnf.infn.it;}
\affiliation{\textsuperscript{$\dagger$}Current address: Italian Space Agency, 00133 Roma, Italy}
%
\date{\today} 
\begin{abstract}
Determining the oxidation states of metals assumes great importance in various applications because a variation in the oxidation number can drastically influence the material properties. As an example, this becomes evident in edible liquids like wine and oil, where a change in the oxidation states of the contained metals can significantly modify both the overall quality and taste. To this end, here we present the MITIQO project, which aims to identify oxidation states of metals in edible liquids utilizing X-ray emission with Bragg spectroscopy. This is achieved using the VOXES crystal spectrometer, developed at INFN National Laboratories of Frascati (LNF), employing mosaic crystal (HAPG) in the Von Hamos configuration. This combination  allow us to work with effective source sizes of up to a few millimeters and improves the typical low efficiency of Bragg spectroscopy, a crucial aspect when studying liquids with low metal concentration. Here we showcase the concept behind MITIQO, for a liquid solution containing oxidized iron. We performed several high-resolution emission spectra measurements,  for the liquid and for different powdered samples containing oxidized and pure iron. By looking at the spectral features of the iron's \kb emission lineshape, we were able to obtain, for a liquid, a result consistent with the oxidized iron powders and successfully quantifying the effect of oxidation.
\end{abstract}
\maketitle
%
\section{Introduction}\noindent
Bragg spectroscopy (BS) is an experimental technique established in the last century to perform ultra-high precision X-ray measurements \cite{haschke2021xray} in fields such as, for example, physics \cite{schnopper1969} and astrophysics \cite{deschamps1986}. In the last three decades, the very low BS efficiency has been improved by the implementation of the mosaic crystals technology \cite{barnsley2003}. Among mosaic crystals, Highly Oriented Pyrolitic Graphite (HOPG) and Highly Annealed Pyrolitic Graphite (HAPG) \cite{grigorieva2019} offer several advantages with respect to normal crystals. They possess an integral reflectivity one order of magnitude higher, an energy range extending above 10 keV, thanks to a smaller lattice parameter (3.514 \AA), as well as the possibility to be shaped in various forms, optimized for specific applications. The Von Hamos (VH) spectrometers \cite{vonhamos1933} provide a further increase in the overall reflectivity, thanks to the sagittal focusing properties of cylindrically bent crystals, opening a wide field of applications \cite{milne2017,rani2020}.\newline
Combining HAPG crystals in the VH configuration with BS, besides increasing the efficiency, is possible to achieve a below 10 eV \cite{seidler2014} energy resolution (FWHM) for X-ray Fluorescence (XRF) measurements. This exceptional capability allows for the precise differentiation of atomic emission lines, providing valuable insights into the elemental composition and properties of materials \cite{debeer2020}.\newline 
In this context the \kb emission line is particularly informative for transition metal elements \cite{glatzel2009}. This emission process occurs in two steps \cite{degroot2001,stohr2023nature}: the creation of a core hole in the 1s orbital and the subsequent filling of the hole with an electron from the 3p state. In contrast to the \ka line, with the emission from the 2p state, can be mainly used to indicate the presence of the metal for elemental analysis, the \kb line can also reveal valuable information about the chemical environment \cite{narbutt1980}, the spin states \cite{pollock2014} and oxidation states \cite{messinger2001} of the metal. The \kb line exhibits a distinct profile, characterized by a prominent \kbt peak and a secondary shoulder known as \kbp, which resides approximately 10 eV lower from the main peak. Notably, the precise position of the \kbp shoulder is highly sensitive to the oxidation state and chemical environment of the metals under investigation, provided that the energy resolution is sufficiently good. By analyzing the \kbp component in relation to the main peak, valuable insights can be extracted regarding the specific oxidation state and chemical surroundings of the involved metal species.\newline
In this specific context, a high-resolution Von Hamos X-ray spectrometer using HAPG mosaic crystals has been developed by the VOXES collaboration at INFN (LNF) \cite{scordo2019,scordo2021,deleo2022}, able to achieve a few eV energy resolution in the soft X-rays range from 2 keV up to 10 of keV also from millimetric sources.\newline
Based on these excellent performances, the MITIQO (Monitoraggio In situ di Tossicità, Indicazione geografica e Qualità di Olio d’oliva, vino e altri liquidi edibili) project aims to conduct high-resolution X-ray spectroscopy of edible liquids by using the VOXES spectrometer to determine the metal's oxidation states. \newline
The identification of metal oxidation states in edible liquids plays a significant role in assessing the quality of wine \cite{tariba2011}. Metals serve as electron sources for redox reactions and hence they act as catalysts in the oxidation process \cite{Pohl2007}. A long-standing problem in winemaking is the browning  of wine \cite{li2008}. Understanding and studying this process is crucial not only for preventing it, but also for finding alternatives to sulphur dioxide, a generally used antioxidant, which may affect the quality of the beverage. One method to assess the oxidative browning effect involves analyzing the oxidation states of metals, such as iron and copper in liquids. These metals act as catalysts for non-enzymatic browning in wine, with multiple pathways related to phenols playing a role \cite{li2008}.\newline
The MITIQO could perform a non-destructive measurement on the wine samples implementing a quantitative analysis with a precise determination  of the metal's oxidation states.\newline
In this work, we introduce the MITIQO concept, for a liquid solution with a high concentration of iron, serving as an initial benchmark for liquids with lower metal concentrations, such as edible liquids. This is done, by performing X-ray emission measurements of several iron-containing powders, with different oxidation states. Our results demonstrate that we can recover the right oxidation state of the liquid solution, consistent with those from oxidized solid samples.\newline
The paper is structured as follows. In Section 2 we describe the experimental setup employed for MITIQO and the preparation of the samples. Section 3 outlines the methodology employed for acquiring and calibrating the spectra of both oxidized and unoxidized samples, together with the spectrum obtained from the liquid sample and the discussion on how oxidized iron samples can be differentiated. Section 4 concludes the paper.
\section{Setup and Methods}\noindent
In this section, we outline the primary characteristics of the experimental apparatus. For a more detailed description of the system and its qualifications, we refer to our previous works \cite{scordo2019,scordo2021,deleo2022,scordo2020}. Furthermore, we describe the procedures involved in preparing solid samples and how we integrate the liquid sample into the setup.\newline
\begin{figure}[]
    \centering
    \includegraphics[width=0.5\textwidth]{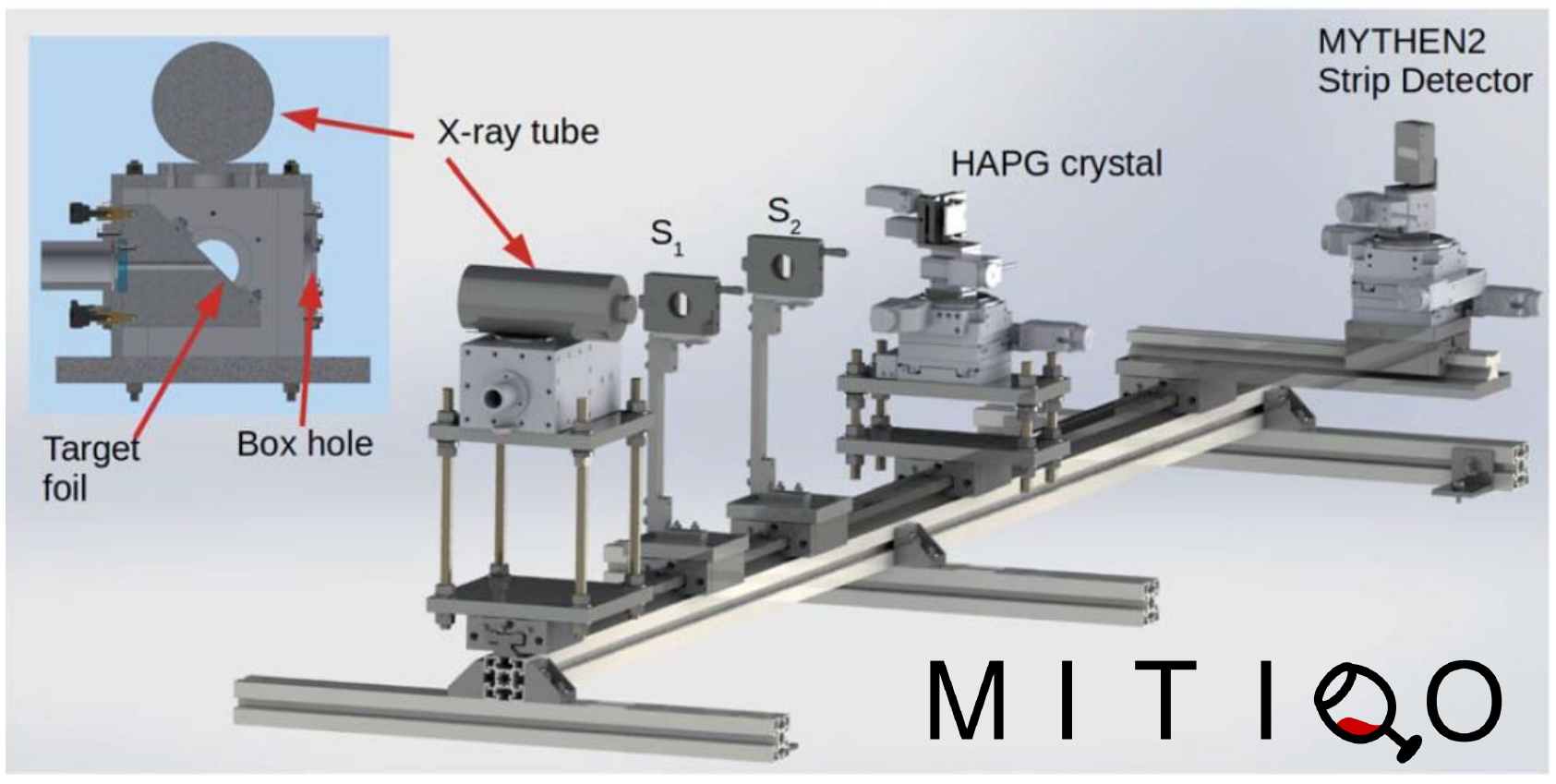}
    \caption{\textbf{Experimental setup for MITIQO.} Overview of the VOXES spectrometer for the MITIQO project at the INFN-LNF \cite{scordo2021}. The setup includes the source box with the X-ray tube, the two slits S\textsubscript{1} and S\textsubscript{2}, the HAPG crystal and the MYTHEN2 detector. The inset in the corner shows the internal view of the source box where the solid sample or the cell for the liquid are placed.}
    \label{fig1}
\end{figure}  
The experimental setup for MITIQO (Figure (\ref{fig1})) consists of an OXFORD XTF-5011 Tungsten X-ray tube, operating at 20 kV voltage and 500 $\mu A$ current, which is used to activate the samples. The emitted fluorescence lines are then shaped by means of two motorized slits (S\textsubscript{1} and S\textsubscript{2}) and reflected by the HAPG crystal onto the surface of the MYTHEN2 strip detector, produced by the DECTRIS company. The HAPG crystal possesses a radius of curvature of $\rho$=206.7 mm, a declared mosaicity of $0.1\pm0.01$, and 100 $\mu$m thickness. The active surface of the MYTHEN2 module measures 32x8 mm\textsuperscript{2} and is equally divided into 640 strips, each with a depth of 450 $\mu$m and 50 $\mu$m pitch. The MYTHEN2 module, slits, and crystal holder are integrated into a remote-controlled motorized system to facilitate precise positioning. Finally, a hole on the back side of the source box allows the usage of a laser for the initial alignment of the system.\newline
The iron-containing samples used in this study are metal salts and they were prepared at Jagiellonian University in Kraków by pouring a measured amount of the sample under study onto a 50 mm wide Kapton tape. Subsequently, another layer of tape was placed on top of the prepared sample for sealing. Finally, the sample was placed under a hydraulic press covered with a thin sponge mat. The samples were compressed with a 50 kN pressure, which allowed to obtain uniform, stable air-free samples. Three different iron powder samples have been prepared: pure iron (>99\%) - Fe; iron(II) sulfate hydrate - \fe{2} (FeSO$_{4}\cdot$7H$_{2}$O) and iron(III) sulfate hydrate - \fe{3} (Fe$_{2}$(SO$_{4}$)$_{3}\cdot$12H$_{2}$O).\newline
To perform the energy calibration we positioned a cobalt foil on top of the iron powders within the mylar. The foil measures 25 mm x 25 mm and has a thickness of 0.125 mm. As we described in the next section, we also utilized a FeCoNi foil (Fe 54\%, Ni 29\%, Co 17\%), of identical dimensions and thickness to the cobalt foil.\newline
The Liquid sample is a medical solution, with the commercial name FerroFolin, used as human iron supplement, containing 40 mg of iron (\fe{3}) in 15 ml solution, for a resulting concentration of 2666 mg/L. The liquid is enclosed in a plastic frame bag, closed on both sides with a 7$\mu$m thick kapton foil. The liquid bag (25x80x2.5) mm$^{3}$ has a total active area of 15x60 mm$^{2}$ to fully cover the effective dimension of the emission source of the target, which was S$_0^\prime$ = 0.8 mm, with S$_0^\prime$ defined as in Ref. \cite{scordo2020}.\newline
\section{Results and Discussion}
\begin{figure}[]
    \centering
    \includegraphics[width=0.5\textwidth]{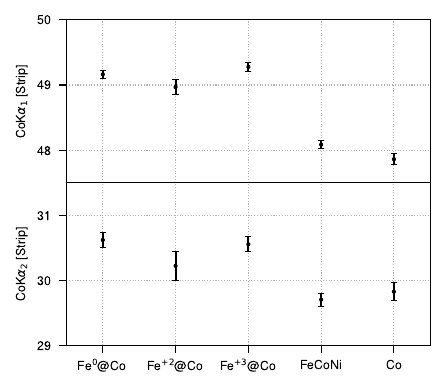}
    \caption{\textbf{Strip positions of cobalt's K$\bm{\alpha}$ lines.} Values of the positions of cobalt's K$\alpha$ lines over the strips for the different iron samples and for the FeCoNi and cobalt foil.}
    \label{fig:copositions}
\end{figure}

\noindent
In what follows, we describe how we obtained the spectra and delineate our approach to detect if iron is oxidized, in both solid and liquid samples.\newline The first step was the energy calibration of the spectra. We needed to compare spectra from different samples and with different oxidation states.\newline
Performing energy calibrations with BS can be challenging due to the complex relationship between the spectrum in space and energy \cite{scordo2020,wansleben2019}. Further difficulties come from the problem under examination. Due to the complex lineshape of the spectrum, as mentioned earlier, only one energy reference, the \kb line of pure iron, is available. Iron's \ka lines can not be used for our calibration, because they are out of the dynamic range of our configuration ($\sim$ 600 eV).\newline
To address this issue, we placed a sample of pure cobalt over the iron powders in the mylar foil, to include the cobalt's \ka lines in the spectrum. This allows the calibration of the pure iron (Fe$^{0}$@Co) sample. Reference values for the K$\beta_{1,3}$ line in oxidized iron are not readily available, as mentioned earlier, because their absolute values are sensitive to the particular sample used. We then performed the measurements for Fe$^{+2}$ and Fe$^{+3}$ with the cobalt foil. We carefully checked that the positions of cobalt lines over the strip of the detector remain unchanged within the uncertainties in the relative spectra (see the first three points of Figure [\ref{fig:copositions}]). This justifies the application of the calibration function of Fe$^{0}$@Co also to the oxidized samples (Fe$^{+2,+3}$@Co).\newline
To extract the position of the peaks we performed a fit, where we approximated the background with a quadratic function. Due to the few eV energy resolution, we used a Voigt profile for all peaks, allowing the Lorentzian smearing factors to vary, while fixing the resolution for all the peaks. In the fit, we included two peaks in the \kb lineshape: the primary peak \kbt and the satellite \kbp between 8 and 14 eV lower from the main one.
The active region where the X-rays are emitted from the sample spans a few millimeters area (i.e. S$_0^\prime$=0.8 mm is the lateral size). This poses a challenge when it is needed to place the cobalt foil on top of the iron sample which can affect the reproducibility of the obtained spectra. To address this issue, we investigated the possibility of extracting the calibration function from an alloy composed of iron, cobalt and nickel. After measuring the spectrum for the FeCoNi sample, we observed a shift in the emission lines of cobalt as compared to Fe$^{0}$@Co (see Figure [\ref{fig:copositions}]). This shift can be attributed to the different vertical position of the cobalt's foil when placed over the iron sample in the mylar. This conclusion gains further support from the result obtained with the pure cobalt foil, which exhibits emission centers similar to those observed in the FeCoNi sample. In the following, we will explore the implications of this shift on the resulting calibration function and it will define a systematic error on the final result. Conversely, there is no shift in the iron's K$\beta$ peaks compared to Fe$^{0}$@Co and FeCoNi, since the iron in both samples has the same vertical position, over the active area where the X-rays are emitted.\newline
\begin{figure}[]
    \centering
    \includegraphics[width=0.5\textwidth]{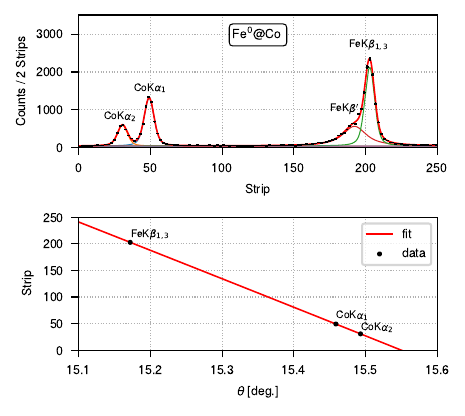}
    \caption{\textbf{Calibration of Fe$^{\mathbf{0}}$@Co.} The uncalibrated spectrum of the Fe$^0$@Co sample is displayed in the top panel, while the corresponding calibration function can be found in the bottom panel. The black line represents the experimental spectrum, which is fitted in red, with the contributions of the individual peaks visualized in various distinct colors.}
    \label{fig:Fe0calib}
\end{figure}  
The calibration function employed in this study, that relates the position along the detector's strip and the angle of reflection $\theta$ (and consequently the energy) is:
\begin{equation} 
\text{Strip}(\theta) = \frac{\sin\theta(A\cot\theta - B)}{\sin(\alpha - \theta)}
\label{eqn:calib}
\end{equation}
This equation is derived from the results presented in \cite{scordo2020}, with ${A,B,\alpha}$ the parameters of the calibration fit. 
The spectrum over the strip and the relative calibration fit for the Fe$^{0}$@Co are shown in Figure [\ref{fig:Fe0calib}]. Reference values were taken from the xraylib library \cite{xraylib2011}, specifically E$_{\text{K}_{\alpha 1}}$ = 6930.3 eV and E$_{\text{K}_{\alpha 2}}$ = 6915.3 eV for cobalt and E$_{\text{K}_{\beta 1,3}}$ = 7058.0 eV for iron. Equation (\ref{eqn:calib}) is then inverted to convert the spectrum over the strips in energy and obtain the calibrated spectra. Additionally, we performed the calibration with FeCoNi, and then compared the results with both calibrations. This comparison allowed us to identify any systematic errors attributable to variations in the vertical positioning of cobalt foil in the Fe$^0$@Co measurements.\newline
Subsequently, we applied both calibration functions to all samples. As an example the spectra for Fe$^{+3}$@Co and the liquid solution are presented in Figure [\ref{fig:spectra}].\newline
The spectrum of Fe$^{3}$@Co exhibits the well-known signatures due to the change of the oxidation state, with the prominent \kbp at lower energies of the \kb. There is a shift towards higher energies in the \kbt compared to the reference value for pure iron. It's worth mentioning that the lines associated with cobalt remain unchanged, as expected. In the case of the liquid spectrum, a similar trend is observed, with an increased background noise due to scattering resulting from the presence of the liquid.\newline 
To investigate the impact of oxidation, we examine the energies of the \kb peaks across different oxidation states. This analysis can be approached by either considering the absolute energy values of the peaks or their energy differences. Absolute differences between \kbt peaks change by approximately $\pm$ 1.6 eV \cite{narbutt1980} with respect to metallic iron, due to the type of bonds present in the surrounding atoms near the metal \cite{debeer2011}. However, since for our spectra, every strip channel corresponds to 1.77 eV (2.55 eV for liquid), it is hard to discern the absolute difference in peak positions across different samples.\newline
Hence, the most effective approach to assess the effect of oxidation lies in analyzing the relative position and shape of peaks in the \kb spectrum, which exhibit a more observable effect. Thus, we computed the energy difference between the peaks as $\Delta \text{E}$ and the ratio of the peaks amplitudes for \kbt and \kbp. In Figure [\ref{fig:oxidation}] the two quantities are shown for different samples and for the liquid solution. 
\begin{figure}[]
    \centering
    \includegraphics[width=0.5\textwidth]{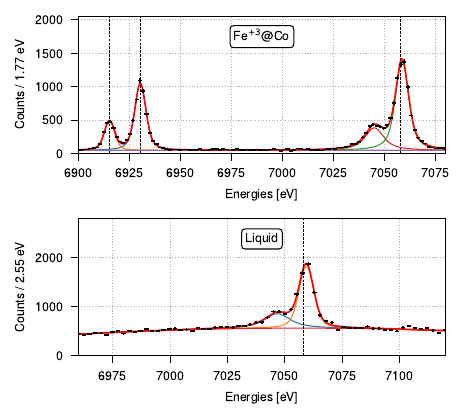}
    \caption{\textbf{Calibrated spectra.} Calibrated spectra for Fe$^{+3}$@Co and liquid solution, where the Fe$^{0}$@Co calibration is used. The legend utilized in Figure (\ref{fig:Fe0calib}) is also employed in all panels and dashed vertical lines represent reference lines for cobalt's K$\alpha_{1,2}$ and pure iron's K$\beta_{1,3}$.}
    \label{fig:spectra}
\end{figure}  
A distinct change of a 4 eV in $\Delta$E can be observed as we move from \fe{0} to \fe{2,3}. It is possible to discern between non- and oxidized states, but it is difficult to distinguish between individual states in the oxidized samples.\newline 
The liquid sample shows a similar trend within the error bars as the oxidized samples, but with larger uncertainties due to a lower iron concentration and lower statistics. 
All quantities were determined using the Fe$^{0}$@Co and FeCoNi calibration functions. By comparing values obtained with different calibrations, we can estimate a systematic error. For instance, in the case of $\Delta$E, we identified a 0.5 eV difference on average across all samples, which is significantly smaller than the impact of oxidation, estimated at approximately 4 eV.
\begin{figure}[]
    \centering
    \includegraphics[width=0.5\textwidth]{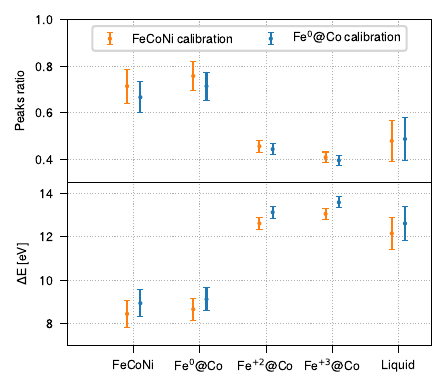}
    \caption{\textbf{Oxidation effect.} Peaks ratio (top panel) and energy difference $\Delta$E (bottom panel) between the peaks of the iron's K$\beta$ across all samples. These values were computed using the calibration functions of FeCoNi (orange) and Fe$^0$@Co (blue).}
    \label{fig:oxidation}
\end{figure}  
%
\section{Conclusions and outlook}\noindent
In this paper, we demonstrated the utility of BS as a technique for differentiating oxidation states of metals. We introduced the VOXES setup, a key ingredient of the MITIQO project, which aims to study the oxidation states of metals in liquids.\newline
For this study, we used pure iron as unoxidized sample and salts of iron and sulfur for the oxidized samples. In addition, we tested our spectrometer with a liquid solution, containing oxidized iron. We discussed the energy calibration for studying the iron's \kb line by including cobalt. We proposed to extract the calibration function from an alloy FeCoNi, and estimated systematic errors of this procedure. Finally, our approach to discern oxidation involved analyzing a specific emission line of iron, the \kb. Two parameters were defined, and we showed that one of them, the difference in energy between the \kbt and \kbp peaks, is an effective discriminator to detect if iron is oxidized or not. 
Furthermore, our results showed that the same indicators for the liquid sample fell within the uncertainties of the oxidized samples, reinforcing the feasibility of applying BS in liquid environments.\newline
As part of future MITIQO plans we aim to detect oxidations in liquid samples with reduced metal concentrations. In edible liquids, such as wine, a metal like iron is typically present at concentrations in the range of mg/L \cite{tariba2011}. Therefore, it is necessary to enhance the efficiency of the spectrometer to operate effectively at such reduced concentrations. Another related aspect involves distinguishing the particular oxidation states (e.g. \fe{2} or \fe{3}). This would require an even better resolution and greater efficiency for detectors, achievable using thinner crystals or high-order of reflection \cite{holzer1997}. Such improvements will allow a more detailed analysis of the \kb lineshape and enable us to compare not only the relative differences but also the absolute positions of the peaks.\newline
As a result of these improvements, it will become possible to improve the characterization and understanding of the oxidation state of metals in edible liquids such as wines and olive oils.\newline
\section{Author contributions}
Conceptualization, M.M. and A.S.; methodology, A.C, M.I., G.M., D.S., F.S., M.M., S.M. and A.S.; software, S.M. and A.C.; formal analysis, S.M.; investigation, M.M.; writing---original draft preparation, S.M., K.P., D.S; writing---review and editing, S.M., K.P., A.P; supervision, A.S. and C.C.; project administration, A.S.; funding acquisition, A.S., R.B. and C.C. All authors have read and agreed to the published version of the manuscript.
\section{Funding}
This research was funded by the MITIQO project n. A0375-2020-36647 from regione Lazio. VOXES was supported by the 5th National Scientific Committee of INFN in the framework of the Young Researcher Grant 2015, no. 17367/2015. This project has received funding from the European Union's Horizon 2020 research and innovation programme EU STRONG-2020, under grant agreement No. 824093.
\section{Data availability}
The data presented in this study are available on request from the corresponding author.
\section{Acknowledgments}
We thank C. Capoccia and G. Fuga from LNF-INFN for their fundamental contribution in designing and building the VOXES spectrometer and Doris Pristauz-Telsnigg, for the support in the preparation of the setup.
\section{Conflicts of interest}
The authors declare no conflict of interest.
%
\vspace{6pt} 
%
\bibliography{references}

\begin{thebibliography}{10}

\bibitem{haschke2021xray}
M.~Haschke, J.~Flock, and M.~Haller, {\em X-ray fluorescence spectroscopy for
  laboratory applications}.
\newblock John Wiley \& Sons, 2021.

\bibitem{schnopper1969}
H.~W. Schnopper and K.~Kalata, ``New high-dispersion high-resolution x-ray
  spectrometer,'' {\em Applied Physics Letters}, vol.~15, no.~5, pp.~134--136,
  1969.

\bibitem{deschamps1986}
J.~Y. Deschamps, R.~Rocchia, and A.~Tarrius, ``A high-resolution bragg
  spectrometer for the observation of the hot interstellar medium x-ray
  emission,'' {\em Astrophysics and Space Science}, vol.~121, pp.~321--332, Apr
  1986.

\bibitem{barnsley2003}
R.~Barnsley, N.~J. Peacock, J.~Dunn, I.~M. Melnick, I.~H. Coffey, J.~A.
  Rainnie, M.~R. Tarbutt, and N.~Nelms, ``Versatile high resolution crystal
  spectrometer with x-ray charge coupled device detector,'' {\em Review of
  Scientific Instruments}, vol.~74, no.~4, pp.~2388--2408, 2003.

\bibitem{grigorieva2019}
I.~Grigorieva, A.~Antonov, and G.~Gudi, ``Graphite optics—current
  opportunities, properties and limits,'' {\em Condensed Matter}, vol.~4,
  no.~1, 2019.

\bibitem{vonhamos1933}
L.~Von~Hámos, ``Röntgenspektroskopie und abbildung mittels gekrümmter
  kristallreflektoren. i. geometrisch-optische betrachtungen,'' {\em Annalen
  der Physik}, vol.~409, no.~6, pp.~716--724, 1933.

\bibitem{milne2017}
J.~Szlachetko, M.~Nachtegaal, D.~Grolimund, G.~Knopp, S.~Peredkov,
  J.~Czapla–Masztafiak, and C.~J. Milne, ``A dispersive inelastic x-ray
  scattering spectrometer for use at x-ray free electron lasers,'' {\em Applied
  Sciences}, vol.~7, no.~9, 2017.

\bibitem{rani2020}
S.~Rani, J.~H. Lee, and Y.~Kim, ``200-mm segmented cylindrical figured crystal
  for von hamos x-ray spectrometer,'' {\em Review of Scientific Instruments},
  vol.~91, no.~1, p.~013101, 2020.

\bibitem{seidler2014}
G.~T. Seidler, D.~R. Mortensen, A.~J. Remesnik, J.~I. Pacold, N.~A. Ball,
  N.~Barry, M.~Styczinski, and O.~R. Hoidn, ``{A laboratory-based hard x-ray
  monochromator for high-resolution x-ray emission spectroscopy and x-ray
  absorption near edge structure measurements},'' {\em Review of Scientific
  Instruments}, vol.~85, 11 2014.
\newblock 113906.

\bibitem{debeer2020}
P.~Zimmermann, S.~Peredkov, P.~M. Abdala, S.~DeBeer, M.~Tromp, C.~Müller, and
  J.~A. {van Bokhoven}, ``Modern x-ray spectroscopy: Xas and xes in the
  laboratory,'' {\em Coordination Chemistry Reviews}, vol.~423, p.~213466,
  2020.

\bibitem{glatzel2009}
P.~Glatzel, G.~Smolentsev, and G.~Bunker, ``The electronic structure in 3d
  transition metal complexes: Can we measure oxidation states?,'' {\em Journal
  of Physics: Conference Series}, vol.~190, p.~012046, nov 2009.

\bibitem{degroot2001}
F.~de~Groot, ``High-resolution x-ray emission and x-ray absorption
  spectroscopy,'' {\em Chemical Reviews}, vol.~101, no.~6, pp.~1779--1808,
  2001.
\newblock PMID: 11709999.

\bibitem{stohr2023nature}
J.~St{\"o}hr, {\em The Nature of X-Rays and Their Interactions with Matter},
  vol.~288.
\newblock Springer Nature, 2023.

\bibitem{narbutt1980}
K.~I. Narbutt, ``X-ray spectra of iron atoms in minerals,'' {\em Physics and
  Chemistry of Minerals}, vol.~5, pp.~285--295, Apr 1980.

\bibitem{pollock2014}
C.~J. Pollock, M.~U. Delgado-Jaime, M.~Atanasov, F.~Neese, and S.~DeBeer,
  ``K$\beta$ mainline x-ray emission spectroscopy as an experimental probe of
  metal–ligand covalency,'' {\em Journal of the American Chemical Society},
  vol.~136, no.~26, pp.~9453--9463, 2014.
\newblock PMID: 24914450.

\bibitem{messinger2001}
J.~Messinger, J.~H. Robblee, U.~Bergmann, C.~Fernandez, P.~Glatzel, H.~Visser,
  R.~M. Cinco, K.~L. McFarlane, E.~Bellacchio, S.~A. Pizarro, S.~P. Cramer,
  K.~Sauer, M.~P. Klein, and V.~K. Yachandra, ``Absence of mn-centered
  oxidation in the s2 $\rightarrow$ s3 transition: Implications for the
  mechanism of photosynthetic water oxidation,'' {\em Journal of the American
  Chemical Society}, vol.~123, no.~32, pp.~7804--7820, 2001.
\newblock PMID: 11493054.

\bibitem{scordo2019}
A.~Scordo, C.~Curceanu, M.~Miliucci, F.~Sirghi, and J.~Zmeskal, ``Pyrolitic
  graphite mosaic crystal thickness and mosaicity optimization for an extended
  source von hamos x-ray spectrometer,'' {\em Condensed Matter}, vol.~4, no.~2,
  2019.

\bibitem{scordo2021}
A.~Scordo, V.~De~Leo, C.~Curceanu, M.~Miliucci, and F.~Sirghi, ``Efficiency
  measurements and simulations of a hapg based von hamos spectrometer for large
  sources,'' {\em J. Anal. At. Spectrom.}, vol.~36, pp.~2485--2491, 2021.

\bibitem{deleo2022}
V.~De~Leo, A.~Scordo, C.~Curceanu, M.~Miliucci, and F.~Sirghi, ``Reflection
  efficiency and spectra resolutions ray-tracing simulations for the voxes hapg
  crystal based von hamos spectrometer,'' {\em Condensed Matter}, vol.~7,
  no.~1, 2022.

\bibitem{tariba2011}
B.~Tariba, ``Metals in wine—impact on wine quality and health outcomes,''
  {\em Biological trace element research}, vol.~144, pp.~143--56, 04 2011.

\bibitem{Pohl2007}
P.~Pohl, ``What do metals tell us about wine?,'' {\em TrAC Trends in Analytical
  Chemistry}, vol.~26, no.~9, pp.~941--949, 2007.

\bibitem{li2008}
H.~Li, A.~Guo, and H.~Wang, ``Mechanisms of oxidative browning of wine,'' {\em
  Food Chemistry}, vol.~108, no.~1, pp.~1--13, 2008.

\bibitem{scordo2020}
A.~Scordo, L.~Breschi, C.~Curceanu, M.~Miliucci, F.~Sirghi, and J.~Zmeskal,
  ``High resolution multielement xrf spectroscopy of extended and diffused
  sources with a graphite mosaic crystal based von hamos spectrometer,'' {\em
  J. Anal. At. Spectrom.}, vol.~35, pp.~155--168, 2020.

\bibitem{wansleben2019}
M.~Wansleben, Y.~Kayser, P.~Hönicke, I.~Holfelder, A.~Wählisch,
  R.~Unterumsberger, and B.~Beckhoff, ``Experimental determination of line
  energies, line widths and relative transition probabilities of the gadolinium
  l x-ray emission spectrum,'' {\em Metrologia}, vol.~56, p.~065007, oct 2019.

\bibitem{xraylib2011}
T.~Schoonjans, A.~Brunetti, B.~Golosio, M.~{Sanchez del Rio}, V.~A. Solé,
  C.~Ferrero, and L.~Vincze, ``The xraylib library for x-ray–matter
  interactions. recent developments,'' {\em Spectrochimica Acta Part B: Atomic
  Spectroscopy}, vol.~66, no.~11, pp.~776--784, 2011.

\bibitem{debeer2011}
M.~A. Beckwith, M.~Roemelt, M.-N. Collomb, C.~DuBoc, T.-C. Weng, U.~Bergmann,
  P.~Glatzel, F.~Neese, and S.~DeBeer, ``Manganese k$\beta$ x-ray emission
  spectroscopy as a probe of metal–ligand interactions,'' {\em Inorganic
  Chemistry}, vol.~50, no.~17, pp.~8397--8409, 2011.
\newblock PMID: 21805960.

\bibitem{holzer1997}
G.~H\"olzer, M.~Fritsch, M.~Deutsch, J.~H\"artwig, and E.~F\"orster,
  ``K${\ensuremath{\alpha}}_{1,2}$ and k${\ensuremath{\beta}}_{1,3}$ x-ray
  emission lines of the $3d$ transition metals,'' {\em Phys. Rev. A}, vol.~56,
  pp.~4554--4568, Dec 1997.

\end{thebibliography}
\bibliographystyle{ieeetr}
\end{document}